\documentclass[aps,superscriptaddress,showpacs,showkeys,nofootinbib, preprint]{revtex4}

\usepackage{graphicx}
\newcommand{\eg}{e.g.,\,}
\newcommand{\ie}{i.e.,\,}
\newcommand{\be}{\begin{equation}}
\newcommand{\ee}{\end{equation}}
\newcommand{\bea}{\begin{eqnarray}}
\newcommand{\eea}{\end{eqnarray}}
\newcommand{\etal}{et al.}

\newcommand{\E}{{\cal E}}
\newcommand{\D}{{\cal D}}
\newcommand{\Poy}{{\cal S}}

\newcommand{\ra}{\rightarrow}

\begin{document}

\bibliographystyle{apsrev}

\title{Cosmology With A Dark Refraction Index}

\author{B. Chen}
\email{Bin.Chen-1@ou.edu}
\affiliation{Homer L.~Dodge Dept.~of Physics and Astronomy, University of
Oklahoma, 440 West Brooks, Rm.~100, Norman, OK 73019, USA}

\author{R. Kantowski}
\email{kantowski@nhn.ou.edu}
\affiliation{Homer L.~Dodge Dept.~of Physics and Astronomy, University of
Oklahoma, 440 West Brooks, Rm.~100, Norman, OK 73019, USA}

\date{\today}

\begin{abstract}
 
We review Gordon's optical metric and the transport equations for the amplitude 
and polarization of a geometrical optics wave traveling in a gravity field. 
We apply the theory to the FLRW cosmologies by associating a refraction 
index  with the cosmic fluid. We then derive an expression for the accumulated effect of a refraction 
index on the distance redshift relations and fit the Hubble curve of current supernova 
observations with a non-accelerating cosmological model.  We also show that some observational effects caused by 
inhomogeneities, \eg the Sachs-Wolfe effect, can be interpreted as being caused by an effective 
index of refraction, and hence this theory could extend to other speed of light communications such as 
gravitational radiation and neutrino fluxes.
\end{abstract}

\maketitle

\section{Introduction}\label{sec:intro}
  The study of the intergalactic medium (IGM) has long been an important field in both astrophysics and cosmology. Current study of the influence of the IGM on distance redshift relation is mainly focused on light absorption, e.g., the magnitudes of supernovae are corrected for light absorption by the IGM when drawing a Hubble diagram. 
We know that light paths in a dielectric medium are different from those in vacuum. According to classical electrodynamics light is altered in both speed and direction (light refraction). The impact of light refraction  on the cosmological distance redshift relation is interesting from both a theoretical and observational point of view.  An interesting and useful theoretical tool to study light refraction in curved spacetime is Gordon's optical metric \citep{Gordon}. The idea of the optical metric is simple, any solution to Maxwell's equations in a curved spacetime filled with a fluid whose electromagnetic properties can be described by a permittivity $\epsilon(x)$ and a permeability $ \mu(x)$ can be found by solving a slightly modified version of Maxwell's equations in a related spacetime with vacuum values for the permittivity and permeability, \ie with $\epsilon(x)=1$ and $\mu(x)=1$. 

Our motive for undertaking this work was two fold. First we wanted to revive the old optical metric theory of \citet{Gordon} and show how one applies it to a modern cosmological model. The Second was to join in current efforts to find alternative explanations of the recently observed cosmic acceleration besides the existence of an exotic $p=-\rho c^2$ material, \ie besides the cosmological constant (see \citep{Marra1, Marra2,  Mattsson, Brouzakis, Conley, Sarkar, Chuang, Alnes, Alnes2, Vanderveld, Chung, Romano, Kolb1}). In Section~\ref{sec:OM}  we introduce the reader to Gordon's optical metric. 
In Section~\ref{sec:GO} we use the WKB approximation to derive the transport equations 
for the amplitude and polarization vector of waves moving through a 
spacetime possessing an optical metric. 
Readers not interested in the mathematical details can skip this section. 
In Section~\ref{sec:FRW} we construct 
the optical metric for Friedmann-Lema\^itre-Robertson-Walker (FLRW) models whose 
cosmological fluid possesses a spatially homogeneous
and isotropic refraction index $n(t)$. We use the optical metric to derive 
both apparent-size, $d_A(z_n)$, and luminosity, $d_L(z_n)$, distance-redshift relations. 
In Section~\ref{sec:SN} we fit the $d_L(z_n)$ of our index of refraction model to 
current supernova data. 
Without a cosmological constant $\Lambda$, we  also estimate the value of $n(t)$ 
required to produce roughly the apparent size distance $d_A(z)$ at last scattering 
required by the position of the acoustic peaks in the angular power spectrum of the 
WMAP data. In Section~\ref{sec:FLAT} we draw our conclusions. 

\section{The Optical Metric}\label{sec:OM}

In GR type theories, a gravity field is described by a metric $g_{ab}$ on a 
four dimensional manifold (we use a +2 signature here). 
We additionally assume the presence of an arbitrarily moving medium with normalized 4-velocity $u^au_a=-1$
 that fills spacetime. For simplicity, we also assume that the  
fluid's  electromagnetic 
properties are linear, isotropic, transparent, and non-dispersive; and can be summarized by 
two scalar functions: a permittivity $\epsilon(x^a)$ and a permeability $\mu(x^a)$.
Following \citet{Ehlers1} in this section, we write the two electromagnetic bivectors as
$F^{ab}(B,E)$ and $H^{ab}(H,D)$. They  satisfy Maxwell's Equations\footnote{Square brackets $[\>\>]$ symbolize complete anti-symmetrization of the enclosed indices.}
\bea\label{Max}
\partial_{[a}F_{bc]}&=&0,\cr
\nabla_bH^{ba}&=&\frac{4\pi}{c}J^a,
\eea
with constitutive relations
\bea\label{constitutive}
H^{ab}u_b&=&\epsilon F^{ab}u_b,\cr
F_{[ab}u_{c]}&=&\mu H_{[ab}u_{c]}.
\eea
(For a familiar example, take Minkowski spacetime with the fluid at rest, $u^a=(1,0,0,0)$.)

The optical metric of \citet{Gordon} is defined as
\be\label{Gordon-Metric}
\bar{g}_{ab}=g_{ab}+(1-{1\over \epsilon\mu})u_au_b,
\ee 
with inverse
\be
\bar{g}^{ab}=g^{ab}+(1-\epsilon\mu)u^au^b.
\label{Gordon-Metric-1}
\ee
The combination $\epsilon\mu$ is related to the usually defined refraction index 
$n(x)\equiv\sqrt{\epsilon\mu}$.
To relate covariant derivatives of the two metrics and to obtain Eq.\,(\ref{Maxbar}) below 
the relationship of the two determinants is needed 
\be
{\rm det}\{\bar{g}_{ab}\}=\frac{1}{\epsilon\mu}{\rm det}\{g_{ab}\}.
\label{det}
\ee
At this point we have one differentiable manifold with two metrics or equivalently two related spacetimes, 
the physical and the optical.  In this paper we are primarily interested in the dynamics of a 
particular type of physical field (radiation) in the optical spacetime. 
All physical objects are described by tensor fields in physical spacetime 
and those of interest here have associated fields in optical spacetime. Where necessary we denote 
optical spacetime fields with a bar.
The optical equivalent of the physical covariant Maxwell field $F_{ab}$ is $F_{ab}$ itself; 
hence the homogeneous Maxwell equations 
are satisfied in both spacetimes, and both share covariant  $4$-potentials.
Using the optical metric the two constitutive equations can be written as a single equation
\be
H^{ab}={1\over\mu} \bar{F}^{ab}\equiv {1\over\mu}\bar{g}^{ac}\bar{g}^{bd}F_{cd},
\label{H}
\ee 
 as can be seen by first expressing the contravariant Maxwell field in optical spacetime as
\be \bar{F}^{ab}=[F^{ab}-(1-\epsilon\mu)u^aF^b_{\>\>\>c}u^c+(1-\epsilon\mu)u^bF^a_{\>\>\>c}u^c],
\ee 
and then contracting separately with $u_b$ and $\epsilon_{abcd}u^c,$ where $\epsilon_{abcd}$ 
is the completely anti-symmetric Levi-Civita symbol. Here the metric dependent bivector 
$\bar{F}^{ab}$ is the metric independent 2-form $F_{ab}$ raised using the optical 
metric Eq.\,(\ref{Gordon-Metric-1}) rather than the physical metric $g_{ab}$.
By using the identity relating the contracted Christoffel symbols to the metric's determinant 
\be
\Bigl\{{\,d\,\atop c\,d}\Bigr\}=\partial_c\log{\sqrt{-\rm{det} \{g_{ab}\}}}\ ,
\label{connection}
\ee
and Eq.\,(\ref{det}), Maxwell's equations (\ref{Max}) can be written using the optical metric as
\bea
\partial_{[a}F_{bc]}&=&0,\cr
\bar{\nabla}_b\left(e^{2\phi}\,\bar{F}^{ba}\right)&=&\frac{4\pi}{c}\bar J^a\equiv\frac{4\pi}{c}\sqrt{\epsilon\mu}\,J^a,
\label{Maxbar}
\eea
where $e^{2\phi}\equiv\sqrt{\epsilon/\mu}$ and the covariant derivatives are taken using the optical metric 
(see \citet{Ehlers1} for details when $J^a=0$).
The form of the inhomogeneous equation is slightly modified and 
hence slightly more complicated,  
\ie the $e^{2\phi}$ term is present in Eq.\,(\ref{Maxbar}); however, 
the advantage is that there are no constitutive equations 
(\ref{constitutive}) to deal with, 
\ie $\bar F^{ab}$ is just $F_{ab}$ raised with the optical metric. Solutions constructed 
in the optical spacetime via Eq.\,(\ref{Maxbar}) directly translate to solutions in 
physical spacetime via Eq.\,(\ref{H}). 
Because the vacuum Maxwell equations are conformally invariant, any metric conformally related to Gordon's can be used to generate $H^{ab}$, see \cite{Ehlers1}.
In the next section we show that  light waves travel along null geodesics at the speed $c$ 
in the optical spacetime (and hence in any conformally related spacetime), 
whereas the corresponding waves travel at speed $c/n$ in physical spacetime.

\section{Geometrical Optics Approximation}\label{sec:GO}

In this section we follow \citet{Sachs0} and \citet{Ehlers2} but use vectors rather than spinors to 
derive the transport equations for the amplitude and polarization of a geometrical optics wave. Readers 
not interested in the  tensor calculus details can skip this section.
 We assume that the electromagnetic wave is planar on a scale large compared with the wavelength, 
but small compared with the curvature radius of spacetime. We write the covariant (and metric independent) 
field tensor as
\be\label{WKB-1}
F_{ab}=\Re\left\{e^{iS/\lambdabar}\left(A_{ab}+{\lambdabar\over i}B_{ab}+O(\lambdabar^2)\right)\right\}, 
\ee 
where $\lambdabar$ is a wavelength related parameter, $S(x^a)$ is the so-called eikonal function and is real,  
and $\Re\{\cdot\}$ stands for the real part. The $A_{ab}$ term represents the geometrical 
optics (GO) approximation and the $B_{ab}$ term is its first order 
correction in both the physical and optical spacetimes. 
Defining the unitless (also metric independent) wave vector $k_a=\partial_a S$ and inserting Eq.\,(\ref{WKB-1}) 
into
the vacuum Maxwell equations ($J^a=0$ in Eq.\,(\ref{Maxbar})) we obtain to order $\lambdabar^{-1}$ 
\bea\label{0th-a}
A_{[ab}k_{c]}&=&0,\cr
\bar{A}^{ab}k_b&=&0,
\eea and to order $\lambdabar^0$
\bea\label{1st-a}
\partial_{[a}A_{bc]}+k_{[a}B_{bc]}&=&0,\cr
\bar{\nabla}_b\bar{A}^{ab}+\bar{B}^{ab}k_b+2\bar{A}^{ab}\phi_{,b}&=&0.
\eea
 All barred contravariant quantities throughout are obtained by raising indices with the optical 
metric, \eg $\bar{A}^{ab}=\bar{g}^{ac}\bar{g}^{bd}A_{cd}$\,;  unbarred are obtained 
by raising with the physical metric $g^{ab}$. 
Equations (\ref{0th-a}) tell us that $\bar{k}^a\equiv \bar{g}^{ab}k_b$ is tangent to 
null geodesics of the optical metric
\bea
\bar{k}^c k_c&=&0,\nonumber\\
\bar{k}^b\bar{\nabla}_b\bar{k}^a&=&0,
\label{Null}
\eea
and that $A_{ab}$ is of the form:
\be
A_{ab}=-2k_{[a}\E_{b]},
\ee  
where $ \E_a$ is spacelike and constrained by $\E_a\bar{k}^a=0$ 
but has the remaining freedom of definition $\E_a\ra \E_a+f(x)k_a$. 
It is $\bar{\E}^a$ that determines the amplitude and polarization of the 
GO wave seen by an observer and it is Eqs.\,(\ref{Null}) that establishes 
the speed of propagation as $c$.

Equation (\ref{1st-a}) tells us that
\be
B_{ab}=2(\E_{[a,b]}-k_{[a}\D_{b]}),
\ee
with a remaining freedom  $\D_a\ra \D_a+g(x)k_a$ and gives as the propagation equation for $\bar{\E}^a$
\be
\dot{\bar{\E}}^a+\bar{\E}^a\theta+\bar{\E}^a\dot{\phi}={\bar{k}^a\over 2}
(\bar{\nabla}_b\bar{\E}^b+k_b\bar{\D}^b+2\phi_{,b}\bar{\E}^b).
\label{Edot}
\ee
The affine parameter derivative symbolized by `\ $\dot{}\ $' is the invariant derivative $\bar{k}^b\bar{\nabla}_b$ 
along the null geodesics 
generated by the vector field $\bar{k}^a$. 
If we now split $\bar{\E}^a$ into a scalar amplitude and a unit polarization vector, \ie
\be
\bar{\E}^a=\E \bar{e}^a, \> \E\geq 0,\> \bar{e}^a\bar{e}_a^*=1,
\ee 
where $*$ means complex conjugate, 
the transport equation for the amplitude $\E$ becomes 
\be
\dot{\E}+\E\theta+\E\dot{\phi}=0,
\label{transport}
\ee 
where $\theta$ is the expansion rate of the null rays defined by the vector field $\bar{k}^a$. 
It is defined by the divergence of $\bar{k}^a$
\be
\theta\equiv{1\over 2}\bar{\nabla}_a\bar{k}^a=\frac{\dot{\sqrt{A}\ \ }}{\sqrt{A}},
\label{theta}
\ee
and is related to the fractional rate of change of the observer independent area $A$ of a small beam of neighboring rays \citep{Sachs0}.
Given $A$, we are able to integrate Eq.\,(\ref{transport})
\be
\E\left(\frac{\epsilon}{\mu}\right)^{1/4}\sqrt{A}=\E_e\left(\frac{\epsilon_e}{\mu_e}\right)^{1/4}\sqrt{A_e}\ ,
\label{transport-solution}
\ee
where the subscript $e$ means evaluate at (or close to) the emitter.

For the calculation at hand we need the amplitude $\E$ only, however, if we were interested in the 
wave's polarization a 
suitable choice for $f(x)$ makes the right hand side of Eq.\,(\ref{Edot}) vanish and 
also makes $\dot{\bar e}^a=0$, \ie 
a particular choice for a polarization vector can be made that is parallelly transported
along the null geodesics of the GO wave.

The GO approximation is just the $O(\lambdabar^0)$ term in Eq.\,(\ref{WKB-1}) \ie
\be
\bar{F}^{ab}=-2\E\,\Re\{e^{iS/\lambdabar}\bar{k}^{[a} \bar{e}^{b]}\}.
\label{Fbar} 
\ee
The frequency and wavelength seen by observers comoving with the fluid can be computed 
using the fact that the phase of the wave changes by $2\pi$ when the 
observer ages by one period of the wave $\tau$, or respectively steps a spatial 
distance of one wavelength $\lambda$ in the direction of the wave, $\hat{k}^a$,
\bea
2\pi&=&-(c\tau\, u^a)\partial_a\left(\frac{S}{\lambdabar}\right)
=-\frac{c\tau}{\lambdabar} (u^ak_a),\nonumber\\
2\pi&=&(\lambda\, \hat{k}^a)\partial_a\left(\frac{S}{\lambdabar}\right)
=-\frac{\lambda}{\lambdabar} n\, (u^ak_a).
\label{tau-lambda}
\eea
The natural choice of the constant parameter $\lambdabar$ is the rationalized wavelength ($\lambda_e/2\pi$) at the emitter. 
This requires the eikonal satisfy $n_e(k_au^a)_e=-1$.

The energy and momentum of this wave as seen by a comoving observer in physical spacetime ($u^ag_{ab}u^b=-1$)
is contained in the Poynting $4$-vector
\be
\Poy^a=-c\,T^a_{\ b}u^b=\frac{c}{4\pi}\left[H^{ac}F_{cb}-\frac{1}{4}\delta^a_bH^{dc}F_{cd}\right]u^b,
\label{Poyn}
\ee
where all quantities in Eq.\,(\ref{Poyn}) are evaluated in the physical spacetime. When evaluated using 
Eqs.\,(\ref{H}) and (\ref{Fbar})
\be
\Poy^a=\frac{c}{4\pi\mu}\left[\bar{F}^{ac}F_{cb}-\frac{1}{4}\delta^a_b\bar{F}^{dc}F_{cd}\right]u^b
=-\frac{c}{8\pi\mu}\E^2\bar{k}^a(k_bu^b).
\ee
In the last equality the oscillations have been averaged over.
The energy density and 3-d Poynting vector seen by observer $u^a$ are respectively
\bea
U&=&-\frac{1}{c}\Poy^au_a=\frac{1}{8\pi\mu}\E^2(\bar{k}^au_a)(k_bu^b)
=\frac{n^2}{8\pi\mu}\E^2(k_au^a)^2,\nonumber\\
\Poy^a_\perp&=&\Poy^a-c\,Uu^a=-\frac{c}{8\pi\mu}\E^2(k_bu^b)\left[\bar{k}^a+n^2(k_cu^c)u^a\right],
\label{U}
\eea
with magnitude
\be
\Poy_\perp\equiv\sqrt{S^a_\perp \Poy_{\perp a}}=
\frac{cn}{8\pi\mu}\E^2(k_au^a)^2=
\Poy_{\perp e}\frac{A_e}{A}\frac{\tau^2_e}{\tau^2}.
\label{poynting}
\ee
In the last equality we 
have eliminated the amplitude $\E$ using Eq.\,(\ref{transport-solution}) 
and continue using a subscript $e$ to represent quantities evaluated near the emitter. 
Equation (\ref{poynting}) simply says that the energy flux varies inversely with the beam's 
area and inversely with the square of the period, even in the presence of an index of refraction. 
We will use this expression in the next section to compute the luminosity 
distance-redshift relation for FLRW 
cosmologies that are filled with a transparent optical material.

\section{Optical Metric for Robertson-Walker (RW) spacetimes}\label{sec:FRW}

The familiar spatially homogeneous and isotropic Robertson-Walker (RW)  metric can be written as
\be
ds^2=-c^2\,dt^2+R^2(t)\left\{{dr^2\over 1-kr^2}+r^2(d\theta^2+\sin^2\theta\, d\phi^2)\right\},
\label{FRW}
\ee 
where $k=1,0,-1$ for a closed, flat or open universe, respectively. The cosmic fluid associated with the RW
metric is at rest in the co-moving spatial coordinates $(r,\theta,\phi)$ and hence has a 
$4$-velocity $u^a=\delta^a_t/c$. We assume that the cosmic fluid has associated with it a homogeneous
and isotropic refraction index which depends only on the cosmological time $t,$ \ie $\sqrt{\epsilon\mu}=n(t)$. 
From Eq.\,(\ref{Gordon-Metric}) only the $\bar{g}_{tt}$ component of the optical metric is seen to differ 
from the physical metric, \ie
\be
\bar{g}_{tt}=-{c^2\over n^2(t)}\,.
\label{FRWbar}
\ee 
The radial null geodesics of the optical metric are found by fixing ($\theta,\phi$) and integrating
\be\label{radial-metric}
\bar{ds}^2=-{c^2\over n^{2}(t)}dt^2+R^2(t){dr^2\over 1-kr^2}=0.
\ee
Two such geodesics traveling between the origin and a fixed comoving point $r$ satisfy
\be
\int^{t_o}_{t_e}{c\,dt\over n(t)R(t)}=
\int^{t_o+\Delta t_o}_{t_e+\Delta t_e}{c\,dt\over n(t)R(t)}=
\int^{r}_{0}{dr\over \sqrt{1-kr^2}}=
{\rm sinn}^{-1}[r],
\label{orbit}
\ee 
where we have defined
\be
{\rm sinn}[r] \equiv \cases{\sin[r] & $k=+1,$\cr r & $k=0,$\cr \sinh[r] & $k=-1$.}
\ee 
In the above equation, ($t_e,t_o$) and ($t_e+\Delta t_e,t_o+\Delta t_o)$ 
represent the emitting and receiving times of the two respective null signals. 
We see from Eq.\,(\ref{orbit}) that the differences in emission and observation times are related by
\be
{\Delta t_o\over n(t_o)R(t_o)}={\Delta t_e\over n(t_e)R(t_e)},
\ee 
and hence the redshift $z_n$ is given by
\be
1+z_n={\Delta t_o\over \Delta t_e}={n(t_o)R(t_o)\over n(t_e)R(t_e)}={n(t_o)\over n(t_e)}(1+z).
\label{redshift}
\ee
We have used $z_n$ as a measure of the observed frequency change but have also kept the usual 
$z$ which measures the wavelength change.  
The tangent to the radial null geodesic ($\bar{k}^ak_a=0$)  can be found directly 
from Eq.\,(\ref{radial-metric}) by a variation
\be
\bar{k}^a=\alpha\left({n\over c\,R},{\sqrt{1-kr^2}\over R^2},0,0\right),
\label{kbarup}
\ee
with covariant components
\be
k_a=\alpha\left(-{c\over nR},{1\over \sqrt{1-kr^2}},0,0\right).
\label{kdown}
\ee
The constant $\alpha$ is arbitrary  and equivalent to the freedom of 
choosing an affine parameter along a null geodesics. 
In the physical metric the light ray has a timelike tangent vector,
\ie
\be
\bar{k}^ag_{ab}\bar{k}^b=n^2(1-n^2)(k_au^a)^2=\alpha^2\left(\frac{1-n^2}{R^2}\right)<0.
\ee

The eikonal $S$ of the GO approximation Eq.\,(\ref{Fbar}) can easily be found for 
this covariant vector field $k_a$ assuming the spherical wave originates from an 
emitter located at $r=0$ 

\be
S(t,r)=\alpha\left(-\int^{t}_{t_e}{c\,dt\over n(t)R(t)}+
{\rm sinn}^{-1}[r]\,\right).
\ee
To relate $\alpha$ and the constant $\lambdabar$ of Eq.\,(\ref{Fbar}) to comoving wave length,  
we  use  Eqs.\,(\ref{tau-lambda}) and (\ref{kdown})  
\be
2\pi=-\frac{c\tau}{\lambdabar} (u^ak_a)
=\frac{\alpha\, c\tau(t)}{\lambdabar n(t)R(t)}
=\frac{\lambda}{\lambdabar_e}\frac{R(t_e)}{R(t)}.
\ee
The last equality results from choosing $\alpha=R(t_e)$ and $\lambdabar=\lambda(t_e)/2\pi$ 
and confirms our interpretation of the conventional $(1+z)=R/R_e$ as the 
wavelength redshift, see Eq.\,(\ref{tau-lambda}).

We are interested in computing the apparent-size and luminosity distances for 
RW models with an index of 
refraction $n(t)$ and must be careful in doing so. The optical metric gives the correct wave 
trajectories, but because it does not measure distances or times correctly, 
densities and rates will be incorrect. Because angles, areas, and redshifts are easier to calculate 
than energy fluxes we start with the apparent size 
distance-redshift $d_A(z_n)$.  We will then compute the luminosity distance  $d_L(z_n)$ 
by using the 3-d Poynting vector of Eq.\,(\ref{U}).
We use the optical metric in the form given in Eq.\,(\ref{FRWbar}) 
with Eq.\,(\ref{FRW}) because the coordinates ($t,r,\theta,\phi$) have direct physical 
interpretations in the RW metric itself. 
For an example, in the local rest frame of the source (observer), 
the proper time interval is $\Delta t_e$ (respectively $\Delta t_o$) 
instead of $\Delta t_e/n(t_e)$ (respectively $\Delta t_o/n(t_o)$), 
and hence the observed shift in periods, $\Delta t_o/\Delta t_e,$ is correctly given by Eq.\,(\ref{redshift}). 
From  Eq.\,(\ref{FRW}) the apparent size distance (also called the angular size distance) 
of a source at coordinates $(t_e,r)$ 
is just 
\be
d_A=r\,R(t_e),
\label{dA}
\ee
as seen by an observer at $(0,t_o)$ where the three coordinates $(r,t_e,t_o)$ 
are constrained by Eq.\,(\ref{orbit}). To give $d_A(z_n)$ we must use Eqs.\,(\ref{orbit})
and (\ref{redshift}) to eliminate $(r,t_e)$ in terms of $(z_n,t_o)$.
We start by using Eq.\,(\ref{redshift}) to change variables in the remaining integral 
of Eq.\,(\ref{orbit}) from 
$t$ to $z_n$.
The following steps are familiar except for the presence of the index of refraction $n(x)$ 
and the two redshift variables $(z_n,z)$. The dynamical equations of Einstein are 
used to change from $t$ to $R$ and then to $(1+z)=R_o/R$.  
\bea\label{r-z}
{\rm sinn}^{-1}(r)&=&\int^{t_o}_{t_e}{cdt\over n(t)R(t)}=\int^{R_0}_{R_e}
{cdR\over n(R)R\,(dR/dt)}\nonumber\\
&=&{c\over R_0H_0}\int^{z(z_n)}_0 \frac{dz}{n(z)h(z)}\,,
\eea
where 
\be
h(z)\equiv
\sqrt{[\Omega_{\Lambda}+\Omega_k(1+z)^2+\Omega_m (1+z)^3+\Omega_r(1+z)^4]},
\ee
and where
\be
\Omega_k\equiv-{c^2k\over H_0^2R_0^2}=1-(\Omega_\Lambda+\Omega_m+\Omega_r).
\ee 
The wavelength redshift $z(z_n)$ as a function of the frequency redshift $z_n$ is found by 
eliminating $t_e$ in Eq.\,(\ref{redshift}).
We refer to solutions to Einstein's equations with a RW symmetry 
as Friedmann-Lema\^itre-Robertson-Walker or simply FLRW cosmologies.  
The three $\Omega$ constants represent, as usual,  current relative amounts of non-interacting gravity sources: 
vacuum, pressureless matter, and radiation energies. 
From Eq.\,(\ref{dA}) we conclude that the apparent size distance-redshift relation 
for a FLRW cosmology with an index of refraction is
\be\label{dA-z}
d_A(z_n)={1\over (1+z(z_n))}{c\over H_0}{1\over \sqrt{|\Omega_k|}}\hbox{sinn}
\left[ \sqrt{|\Omega_k|}\int^{z(z_n)}_0\frac{dz}{n(z)h(z)}\right].
\ee

To derive the luminosity-redshift relation $d_L(z_n)$ knowing $d_A(z_n)$ one ordinarily uses 
Etherington's \citep{Etherington} result
\be
d_L(z)=(1+z)^2d_A(z).
\label{dLdA}
\ee
If this result were correct with an index of refraction present, 
one would need to know which redshift to use, frequency $z_n$ or wavelength $z$.
To know what to choose we evaluate  the magnitude of the Poynting vector Eq.\,(\ref{poynting}) 
and arrive at the correct replacement for Eq.\,(\ref{dLdA}).  
To find the needed area $A$ we evaluate the expansion $\theta$  of Eq.\,(\ref{theta}) 
using  Eqs.\,(\ref{kbarup}), (\ref{FRW}), (\ref{FRWbar}), and (\ref{det}). We find a simple result
\be
\theta=\frac{\dot{(rR)}}{(rR)}\,,
\ee
and hence the beam area $A\propto rR$ 
from which we have the needed Poynting vector magnitude, Eq.\,(\ref{poynting}),
\be
\Poy_\perp=
\Poy_{\perp e}\frac{4\pi (rR)^2_e}{4\pi(rR)^2}\frac{\tau^2_e}{\tau^2}
=\frac{L_e}{4\pi(rR)^2(1+z_n)^2}
=\frac{L_e}{4\pi d_L^2}.
\label{poynting2}
\ee
The latter identity defines the luminosity distance $d_L$ in terms of 
the total power radiated at the emitter $L_e$ and the flux received, \ie the Poynting vector at the observer,
\be
d_L=rR(1+z_n)=rR_e(1+z)(1+z_n),
\ee
which agrees with the Etherington result Eq.\,(\ref{dLdA}) only if we use one frequency 
redshift factor $(1+z_n)$ from Eq.\,(\ref{redshift}) and one wavelength redshift $(1+z)$. 
Equation (\ref{poynting2}) also confirms a conserved photon number interpretation of the radiation even 
in the presence of a time dependent index of refraction (which has the potential of taking energy out of 
the radiation field). It is equivalent to having a fixed number of photons  
emitted in a time $\Delta t_e$ each having energy $h\nu_e$ 
and all being collected over an area $4\pi (r R)^2$ in a time $\Delta t_o$ but with redshifted energy $h\nu_o$.

\section{An Effective Index of Refraction Induced by the Sachs-Wolfe effect}\label{sec:SW}

Up to now, we have been considering a refractive index modeled after the one generated by induced 
electromagnetic polarizations in inter and intra galactic media, dark or otherwise. 
Lensing has long been interpreted as a
gravitationally induced refraction effect, and here we suggest that to $1^{st}$ 
order, inhomogeneities in the flat FLRW models 
are equivalent to effective indices of refraction. 
 
\citet*{Sachs} were two of the first to consider the effect of perturbations of 
the homogeneous and isotropic models on optical observations. 
In that classic paper, the authors used perturbations in the flat, \ie $k=0,$ FLRW 
spacetime to study the angular fluctuations in the CMB.  
They used a conformally flat version of the metric
\be
ds^2=R^2(\eta)\left[\eta_{ab}+h_{ab}\right]dx^adx^b
\label{SW-start}
\ee 
with dimensionless coordinates and worked in a comoving gauge to derive the equations 
governing the evolution of the metric perturbation $h_{ab}$ and perturbations of 
the energy-momentum tensor $\delta T_{ab}.$ 
Here the 
 conformal time coordinate of the flat Minkowski metric $\eta_{ab}$ is  
$x^0=\eta$ and for the pressureless case is familiarly related to the comoving FLRW time coordinate $t$ by 
$\eta=(3\,tH_0/2)^{1/3}$.
The three Euclidean spatial coordinates are labeled by 
letters of the Greek alphabet. 
They then considered the deviations of null geodesics 
from the unperturbed case and derived 
the now famous temperature fluctuations in the micro-wave background caused 
by $h_{ab}$, 
see Eq.\,(42) of Ref.\,\citep{Sachs}.  
Among the scalar, vector, and tensor perturbation 
modes in dust ($p=\delta p=0$, see Eq.\,(22) of 
Ref.\,\citep{Sachs}), the scalar density perturbations, \ie the relatively decreasing $A(x^\gamma)$ mode 
and relatively increasing $B(x^\gamma)$ mode  
(responsible for the famous Sachs-Wolfe effect)  give
\be
h_{\alpha\beta}=-\frac{1}{\eta^3}A_{,\alpha\beta}+\delta_{\alpha\beta}B+\frac{\eta^2}{10}B_{,\alpha\beta}\,,
\ee
and $h_{0a}=0$. The arbitrarily specified form of the scalar modes 
are related to the density perturbation $\delta\rho$ through  
Poisson's equation  
\be
\delta\rho={H^2_0\over 32\pi G}\nabla^2\left(\frac{6A}{\eta^9}-\frac{3B}{5\eta^4}\right),
\ee 
(See Eq.\,(22) of Ref.\,\citep{Sachs}).
The time component of the perturbation $h_{00}$ vanishes because 
of a comoving gauge choice ($u^a\propto \delta^a_0$) and $h_{0\alpha}$ 
is only present for the rotational perturbations. 
To connect this metric to Gordon's optical metric 
we must make the following non-gauge change of coordinates 
\bea
x^\alpha&=&\bar{x}^\alpha+\frac{1}{2\bar{\eta}^3}A_{,\alpha}(\bar{x})
-\frac{\bar{\eta}^2}{20}B_{,\alpha}(\bar{x}),\nonumber\\
\eta&=&\bar{\eta}\left(1-\frac{3}{2\bar{\eta}^5}A(\bar{x})-\frac{1}{10}B(\bar{x})\right),
\label{non-gauge-trans}
\eea
and obtain to 1$^{st}$ order
\be
ds^2=R^2(\bar{\eta})\left(1-\frac{3}{\bar{\eta}^5}A(\bar{x})+\frac{3}{10}B(\bar{x})\right)^2
\left\{
-\frac{d\bar{\eta}^2}{\left(1-\frac{6}{\bar{\eta}^5}A(\bar{x})+\frac{3}{5}B(\bar{x})\right)^2}
+d\bar{r}^2+\bar{r}^2(d\theta^2+\sin^2\theta d\phi^2)
\right\}.
\label{SW-Gordon}
\ee
This simply says that the Sachs-Wolfe metric is a conformally 
transformed Gordon metric corresponding to a k=0, FLRW metric with a 
spacetime index of refraction $n=1-6A(\bar{x})/\bar{\eta}^5+3B(\bar{x})/5$. 
Since conformal transformations don't alter light cones (see \cite{Ehlers1}) 
we have arrived at the connection of 
null geodesics of Sachs-Wolfe's density perturbations with the light curves of an unperturbed 
FLRW spacetime possessing an index of refraction. In contrast to the 
homogeneous optical fluid discussed in Section \ref{sec:FRW}, the comoving 
frame of the index of refraction in Eq.\,(\ref{SW-Gordon}) is not the same as the
comoving frame of the matter density in Eq.\,(\ref{SW-start}). However, they are related by the 
coordinate change of Eq.\,(\ref{non-gauge-trans}).

We have our doubts about extending the index 
of refraction comparison beyond linear perturbations, and 
make no claims as to that possibility. Such an extension 
would be quite interesting  because old work 
\citep{Kantowski1, Kantowski2} on non-linear 
observational effects in Swiss Cheese cosmologies 
are again in the literature  
\citep{Marra1, Marra2,  Mattsson, Brouzakis} 
also hoping to find sources of apparent acceleration 
other than a cosmological constant.  Work on interpreting  
effects of local density perturbations on the Hubble curve 
are numerous 
\citep{Conley, Sarkar, Chuang, Alnes, Alnes2, Vanderveld, Chung, Romano, Kolb1},  
the results of which can be compared to the above in the $1^{st}$ order regime.

\section{Fitting Supernova Data With A Refraction Index Model}\label{sec:SN} 
 
\begin{figure*}
\includegraphics{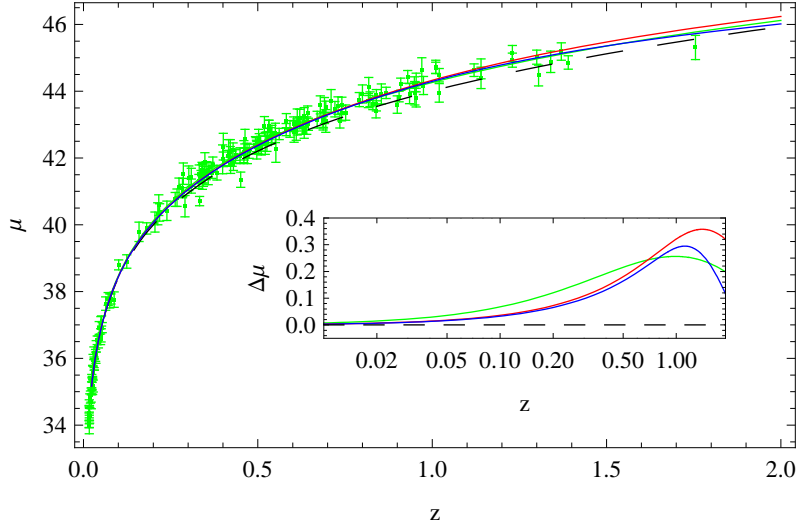}
\caption{Distance modulus $\mu$ versus redshift $z.$ Dashed curve:  
$\Omega_\Lambda=0,\Omega_m=0.3, n=1;$ 
Green curve: $\Omega_\Lambda=0.7,\>\Omega_m=0.3,\> n=1;$ 
Red curve: $\Omega_\Lambda=0,\>\Omega_m=0.04,\> n(z)=1+0.1z^2-0.045z^3;$ Blue curve:  $\Omega_\Lambda=0,\>\Omega_m=0.3,\> n(z)=1+0.15z^2-0.05z^3.$ $\Delta\mu(z)$ are given in the inset. The difference is taken with respect to the fiducial case where $\Omega_\Lambda=0,\Omega_m=0.3, n=1.$   
\label{fig:Mu-Z-Plot}}
\end{figure*}

\begin{figure*}
\includegraphics{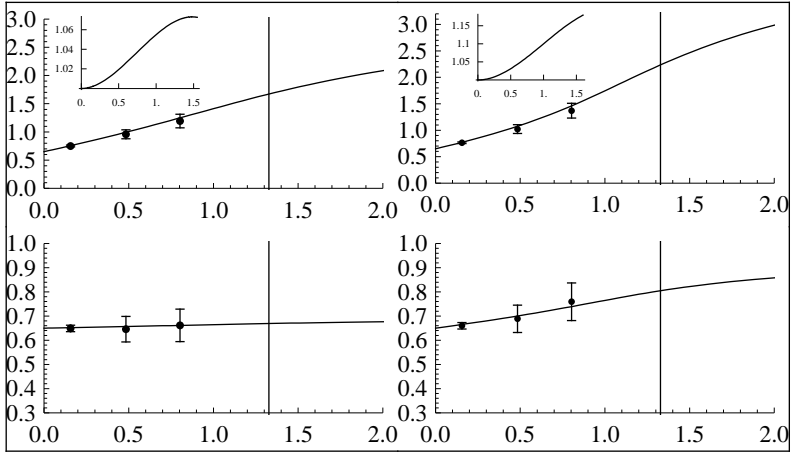}
\caption{$H(z)$ (upper panel) and $\dot{R}(z)/R_o$ (lower panel) curves for the two $n\neq 1$ models. Left column parameters:  $\Omega_\Lambda=0,\>\Omega_m=0.04,\> n(z)=1+0.1z^2-0.045z^3;$ Right column parameters: $\Omega_\Lambda=0,\>\Omega_m=0.3,\> n(z)=1+0.15z^2-0.05z^3.$  
\label{fig:H-Z-Plot}}
\end{figure*}

In this section we use the index of refraction model of Section~\ref{sec:FRW} to
fit the current supernova data \citep{Riess2,Riess,Astier,Davis, Wood}.  
We use the $178$ supernova from the gold sample
\footnote{http://braeburn.pha.jhu.edu/$\sim$ariess/R06/.} 
with redshifts greater than $cz = 7000\> \hbox{km}/\hbox{s}$ , see Fig.\,\ref{fig:Mu-Z-Plot}. 
The Hubble constant we use is $H_0=65$  km/s/Mpc
and since we are concerned with the matter dominated era, we exclude radiation ($\Omega_r=0).$
We compare the distance modulus versus redshift, $\mu(z)$, 
of  the concordance model, $\Omega_\Lambda=0.7,\>\Omega_m=0.3,\> n=1$ with 
two $n(z) > 1$ models. The first is a baryonic matter only  model ($\Omega_m=0.04$)  
and no cosmological constant ($\Omega_\Lambda=0$) with $n(z)=1+0.1z^2-0.045z^3.$ 
The second model includes a dark matter contribution, 
$\Omega_m=0.3,$ no cosmological constant, and $n(z)=1+0.15z^2-0.05z^3.$ 
Also included is a now disfavored dark matter only  model, 
$\Omega_\Lambda=0,\>\Omega_m=0.3,\>n=1.$ In the inset of 
Fig.\,\ref{fig:Mu-Z-Plot}, we use this case  
 to compare with the two $n\neq 1$ models and with the concordance model. The critical 
redshift region is between 
$0.2<z<1.2,$ where most of the supernova data is concentrated.   
Both $n\neq 1$ models fit the data much better in this region than 
models with the same $\Omega$ parameters but with no refraction. 
The two refraction indices are plotted in the insets of 
Fig.\,\ref{fig:H-Z-Plot}. As the reader can easily see the 
effects of a suitable index of refraction $n(z)$ can 
simulate the accelerating effects of a cosmological constant.

The Supernova data is currently considered the best evidence for the existence 
of dark energy because of the observed acceleration in 
the expansion of the universe (see Fig.\,7 of Ref.\,\citep{Riess}). 
For the homogeneous FLRW models, the acceleration is directly related to  
density and pressure by
\be
{\ddot{R}\over R}=-{4\pi G\over 3}(\rho+3\frac{p}{c^2}).
\label{Einstein}
\ee 
A true observed acceleration, \ie $\ddot{R}>0,$ requires 
 $p<-\rho c^2/3,$ and hence implies an unusual equation of state such as vacuum 
energy ($p=-\rho c^2).$  What we show here is that an overlooked index of refraction 
can cause a misinterpretation of the Hubble curve, suggesting an acceleration.
In Fig.\,\ref{fig:H-Z-Plot}  we plot $H(z)$ 
and $\dot{R}(z)/R_o=H(z)/(1+z)$ for the two $n \neq 1$ models. They can be compared with similar plots in Ref.\,\citep{Riess}.
The data points are plotted using flux averaging \citep{Wang0,Wang1} and uncorrelated redshift binning 
\citep{Wang2} algorithms. To apply these techniques to non-flat cases,
 we define 
\bea
g(z_n)&\equiv&\int_0^{z(z_n)}\frac{dz}{n(z)h(z)}\cr
&=&{1\over \sqrt{|\Omega_k|}}\hbox{sinn}^{-1}\left[{\sqrt{|\Omega_k|}\over 1+z_n}{H_0\over c}
10^{{\mu\over 5}-5}\right],
\label{g}
\eea 
where $\mu$ is the distance modulus
\be
\mu=5\log {d_L\over 1\>\hbox{Mpc}}+25.
\ee We furthermore defined
\be
x_i={g(z_n^{i+1})-g(z_n^i)\over z^{i+1}(z_n^{i+1})-z^i(z_n^i)},
\ee 
which when averaged 
 inside each bin gives us an estimate of the 
inverse of the product of $n(z)$ with $h(z)\equiv H(z)/H_0$ 
(compare with Eq.\,(5) of Ref.\,\citep{Wang2}). The presence of an index of 
refraction produces a degeneracy in determining the value of $H(z)$ and hence in $\ddot{R}(t)$.
A suitably decreasing $n(t)$ and a non-accelerating $R(t)$ will mimic an accelerating universe.

We can see that our index of refraction models fit the data well. However, 
we need to remind the reader that the binned data plotted on the $H(z)$ and $\dot{R}(z)/R_0$ curves 
are model dependent. The binning process as designed in Ref.\,\cite{Wang2} 
requires use of $d_L(x_n)$, \ie $g(z_n)$ in Eq.\,(\ref{g}). Rather than using this technique
to argue for an observed accelerating $H(z)$, we argue for an observed $n(z)$ 
with a non-accelerating $H(z)$. The point we make is that we can fit the $\mu(z)$ data with no 
$\Lambda$, and are able to get rid of the acceleration. 

\section{Flatness of the Universe}\label{sec:FLAT}

The conclusion drawn from the latest WMAP data, when combined with the SNe Ia Hubble curve,  
that vacuum energy exists, depends crucially on many unconfirmed theoretical assumptions 
including the adiabatic power law assumption 
for the initial perturbation spectrum \citep{Spergel, Spergel2}. 
Such observations have motivated efforts to find 
ways to produce a perceived acceleration other than by 
a real $\Lambda$ \citep{Conley, Sarkar, Chuang, Alnes, Alnes2, Vanderveld, Chung, Romano, Kolb1}. 
 This section is another such effort.

\begin{figure*}
\includegraphics{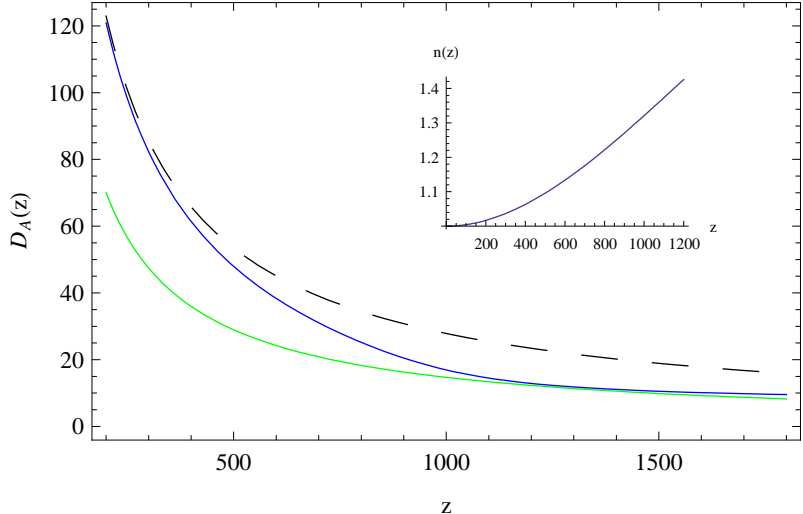}
\caption{Dashed curve: $\Omega_\Lambda=0,\>\Omega_m=0.3,\> n(z)=1;$  Blue Curve:  $\Omega_\Lambda=0, \> \Omega_m=0.3,\>  n(z)=1+4.45\times 10^{-7}z^2-1.25\times10^{-10}z^3;$ Green curve: $\Omega_\Lambda=0.7,\>\Omega_m=0.3,\> n=1.$  \label{fig:NzBAO}}
\end{figure*}

The angular position of the first acoustic peak is commonly believed to be the 
strongest piece of evidence for the flatness of the universe. 
The characteristic wavelengths of the acoustic oscillations at the last 
scattering surface depend very weakly on $\Lambda$, but their observed angular size
as seen 
by us now depends significantly on a combination of  
$\Omega_m$ and $\Omega_\Lambda$, see Eq.\,(\ref{dA-z}).
Assuming  our Universe is of the FLRW type with no refraction index, 
a first acoustic peak at $\sim 0.8^{\rm o}$ is almost  fit by a flat universe.\footnote{
Within the context of a power law 
$\Lambda$CDM model ($w=1$), WMAP data alone does not rule out non-flat 
models. With a prior on the Hubble constant, 
$H_0>40\>\hbox{km}\hbox{s}^{-1}\hbox{Mpc}^{-1},$ or combined 
with other astronomical observations, such as SDSS LRG sample, 
HST constraint on the Hubble constant, or SNe data, WMAP data strongly favors a nearly 
flat universe with nonzero vacuum energy, see Table 12, Fig.\,20, 
and Fig.\,21 of \citep{Spergel2}.  For a more general model of dark 
energy, e.g., one with a time evolving equation of state parameter $w\neq 1$ 
instead of a cosmological constant $\Lambda,$ significant spatial 
curvature is still allowed even when $H_0$ is not restricted to be 
small, see e.g. \citep{Ichikawa-1,Ichikawa-2}. 
}
With a suitable index of refraction and no cosmological constant 
we can produce an  angular 
diameter distance comparable to the angular diameter distance of a flat cosmology at 
any given redshift, independent of the Hubble parameter $H_0.$ In Fig.\,\ref{fig:NzBAO} we have used a
$\Omega_\Lambda=0,\>\Omega_m=0.3$ model with an index of refraction  $n(z)$ 
shown in the inset. We chose this $n(z)$ because it produces similar 
distances over the large redshift range $z> 1000$. 
To conclude that WMAP implies flatness requires the acceptance of the accuracy of 
theoretical assumptions beyond the initial perturbation spectrum; \eg even the accuracy 
of the optics of homogeneous FLRW models is now being questioned as was pointed out in Section~\ref{sec:SW}.

\section{Discussion}

We have reviewed Gordon's optical metric theory \cite{Gordon} 
which incorporates an index of refraction into its geometry. We then used the optical spacetime 
to derive the transport equations for the amplitude and polarization of a geometrical optics 
wave. We applied it to the homogeneous, $\Lambda=0$, FLRW models and estimated the 
refraction index needed to fit current SNe Ia and WMAP data. We found that an $n(z)\approx 1.07$  
 at redshift $z=1.5$ in a baryon only model, or an $n(z)\approx 1.15$ at $z=1.5$ in a dark matter model, 
could easily fit the supernova data (see Fig.\,\ref{fig:Mu-Z-Plot}, and Fig.\,\ref{fig:H-Z-Plot}), 
and that an $n(z)$ as big as $1.3$ at the last scattering surface in a dark matter model 
would give the same angular diameter distance $d_A(z)$ 
as the concordance model (see Fig.\,\ref{fig:NzBAO}). 

The question is, where could such an index of refraction come from?
If it had its origin in atomic dipole moments or charges in plasmas the densities
 would have to be much larger than they actually are. A critical mass density now is about 
$8\times 10^{-30}\>\hbox{g}\cdot \hbox{cm}^{-3}$, which translates to 
$1\times 10^{-20}\> \hbox{g}\cdot \hbox{cm}^{-3}$ at $z=1100$. 
The density of air on the earth is about $1.2\times 10^{-3}\>\hbox{g}\cdot \hbox{cm}^{-3},$ 
some $10^{17}$ times denser  than the universe at recombination and 
yet its index of refraction is only $n=1.0003$. We conclude that there is little hope for a
baryon-lepton origin for $n$. 
A long shot would be a colorless index of refraction for the mysterious dark matter. 
 
Severe limits have already been estimated on direct interactions of the dark matter particles
with photons caused by fractional charge \cite{Taoso, Davidson} ($q/e<10^{-5}-10^{-7}$ depending 
on the particle's mass) and by electric/magnetic 
dipole moments \cite{Sigurd} (dipole moment $< 3\times 10^{-16}e$\,cm).
General limits on photon interaction rates have even been estimated by requiring the associated 
collisional damping scale be small enough to allow structures larger than $100\,\hbox{kpc}$ to form \cite{Boehm}.
Proposing an $n$ of unknown source is perhaps outlandish, but not much more than proposing a 
non-intuitive repulsive cosmological constant 
$\Lambda$ to produce acceleration. 
Even though the latter has become fashionable, we wish to add a dark 
index of refraction theory to the lists of alternatives to think about.

In this paper we developed the  general framework needed for using Gordon's 
optical metric in cosmological observations, but have 
applied it only to an index of refraction model which is homogeneous and isotropic.
If the dark matter and it's assumed  index of  refraction were truly homogeneous 
we could have additionally proposed a redshift dependence 
for $n(z)$ modeled after a dilute dielectric gas or plasma. 
However, such a model would still contain unknowns equivalent to 
ionization densities and/or molecular polarizabilities. 
Instead we chose a phenomenological expression in the form of a 
cubic containing two parameters which we adjusted (\ie $n(z)=1+\alpha z^2+\beta z^3$). 
Such a simple starting point is prudent because we know the real universe is filled with  
low density voids, and high density condensations, as well as associated velocity perturbations
all of which would modify the
refraction index $n(x^a)$. If an index of refraction model such as the one proposed here has merit, 
future efforts can look into how such perturbations,
including local variations in the magnetic field 
of the intergalactic medium, might impact distance-redshift.
However, the optical metric 
theory Eq.\,(\ref{Gordon-Metric}) is still the applicable theory.  
We also leave to the future, further exploration of  the equivalence
of the optical effects of gravitational inhomogeneities (beyond the linear perturbation results 
of Sachs-Wolfe in Section~\ref{sec:SW}) and
our index of refraction proposal. Complete equivalence would be 
quite interesting and useful in light of the current interest in 
Swiss Cheese optics \citep{Kantowski1, Kantowski2, Marra1, Marra2,  Mattsson, Brouzakis}. 
Modifications in distance-redshift caused by random spacetime 
perturbations could then be interpreted as being caused by an effective index of refraction. 
The idea of an optical metric can also be applied to other massless particles which 
follow null geodesics in vacuum.  
If the presence of material causes interference of the propagating waves 
of the particles, an effective refraction index should exist. 
For an example, gravitons and some neutrinos are massless and 
local inhomogeneities, such as the Sachs-Wolfe perturbations 
discussed in Section~\ref{sec:SW}, would alter propagation of their waves \citep{Zeldovich}. 

The idea of solving cosmological problems via a changing light speed model 
is not new, see \citep{Albrecht, Barrow, Bassett}.   Ellis \cite{Ellis1} 
has recently pointed out consistency constraints required of such theories. 
Our proposal is  fundamentally
different from those cited above in that it is based on a classical 
electrodynamics analogy (the cosmological fluid simply has an unexpected 
refraction index which reduces the propagation speed of electromagnetic waves). 
Because we are not proposing a change in the vacuum light speed 
$c$ or the limiting causal speed, the proposal is not subject to Ellis's criticism. 

 Finally we note that since an accelerating universe is consistent with other
observations, such as Baryon Acoustic Peaks detected in galaxy surveys \citep{Tegmark, Tegmark2, Eisenstein} and the interesting $H(z)$ relation obtained from ages of passively evolving galaxies in \cite{Simon}, 
additional comparisons with refraction models are in order.

\section{Acknowledgments}

This work was supported in part by NSF grant AST-0707704, and US DOE
Grant DE-FG02-07ER41517.

\label{lastpage}

\end{document}